\documentclass{aa}

\usepackage{graphicx}

\newcommand{\xmm}{{\it XMM-Newton}}

\newcommand{\kepler}{{\it Kepler}}

\newcommand{\ginga}{{\it Ginga}}
\newcommand{\rxte}{{\it RXTE}}

\begin{document}

\title{Similar shot profile morphology of fast variability in cataclysmic variable, X-ray binary and blazar; the MV\,Lyr case}
\titlerunning{Shot profile in MV\,Lyr vs Cyg\,X-1 and blazar}
\authorrunning{Dobrotka et al.}

\author{A.~Dobrotka \inst {1}, H. Negoro \inst {2} and S. Mineshige \inst {3}}

\offprints{A.~Dobrotka, \email{andrej.dobrotka@stuba.sk}}

\institute{Advanced Technologies Research Institute, Faculty of Materials Science and Technology in Trnava, Slovak University of Technology in Bratislava, Bottova 25, 917 24 Trnava, Slovakia
\and
			Department of Physics, Nihon University, 1-8 Kanda-Surugadai, Chiyoda-ku, Tokyo 101-8308, Japan
\and
			Department of Astronomy, Graduate School of Science, Kyoto University, Sakyo-ku, Kyoto 606-8502, Japan
}

\date{Received / Accepted}

\abstract
{The cataclysmic variable MV\,Lyr was present in the \kepler\ field yielding a light curve with the duration of almost 1500 days with 60 second cadence. Such high quality data of this nova-like system with obvious fast optical variability show multicomponent power density spectra as shown previously by Scaringi et al.}
{Our goal is to study the light curve from different point of view, and perform a shot profile analysis. We search for characteristics not discovered with standard power density spectrum based methods.}
{The shot profile method identifies individual shots in the light curve, and averages them in order to get all substructures with typical time scales. We also tested the robustness of our analysis using simple shot noise model. Although, the principle of this method is not totally physically correct, we use it as a purely phenomenological approach.}
{We obtained mean profiles with multicomponent features. The shot profile method distinguishes substructures with similar time scales which appear as a single degenerate feature in power density spectra. Furthermore, this method yields the identification of another high frequency component in the power density spectra of \kepler\ and \xmm\ data not detected so far. Moreover, we found side-lobes accompanied with the central spike, making the profile very similar to another \kepler\ data of blazar W2R\,1926+42, and \ginga\ data of Cyg\,X-1. All three objects show similar time scale ratios of the rising vs. declining part of the central spikes, while the two binaries have also similar rising profiles of the shots described by a power-law function.}
{The similarity of both binary shot profiles suggests that the shots originate from the same origin, e.g. aperiodic mass accretion in the accretion disc. Moreover, the similarity with the blazar may imply that the ejection fluctuations in the blazar jet are connected to accretion fluctuations driving the variability in binaries. This points out to connection between jet and the accretion disc.}

\keywords{accretion, accretion discs - stars: novae, cataclysmic variables - stars: individual: MV\,Lyr - X-rays: binaries - galaxies: BL Lacertae objects: general}

\maketitle

\section{Introduction}
\label{introduction}

Several kinds of objects in the universe are powered by accretion, ranging from small binary systems as cataclysmic variables (CVs), through symbiotic (SSs) and X-ray binaries (XRBs) to huge active galactic nuclei (AGNs). Usually the accretion process generates an accretion disc\footnote{If not prevented by strong magnetic field of the compact object.} around a central compact object ranging from white dwarf in CVs or SSs, through main sequence stars in SSs and stellar black holes in XRBs to supermassive black holes in AGNs. The common accretion process generating very similar radiation characteristics makes these objects a very suitable target to study the physics of accretion in many conditions and on very large interval of time scales.

The existence of the accretion process is usually seen as fast variability (a.k.a. flickering) in all mentioned objects (see, e.g. \citealt{mchardy1988}, \citealt{miyamoto1992}, \citealt{bruch2015}, \citealt{vaughan2003a}). Such flickering has three basic observational characteristics; 1) linear correlation between variability amplitude and log-normally distributed flux (so called rms-flux relation) observed in all variety of accreting systems such as XRBs or AGNs (\citealt{uttley2005}), CVs (\citealt{scaringi2012b}, \citealt{vandesande2015}) and SSs (\citealt{zamanov2015}), 2) time lags where flares reach their maxima slightly earlier in the blue than in the red (\citealt{scaringi2013}, \citealt{bruch2015}) and 3) red noise or band limited noise with characteristic break frequencies in power density spectra (PDS, see e.g. \citealt{sunyaev2000}, \citealt{scaringi2012a}, \citealt{dobrotka2014}, \citealt{dobrotka2015b}).

A PDS technique usually generalizes the available information in the light curve, and additional knowledge about the flickering nature can be studied from the profile of the flickering flares. \citet{negoro1994} proposed such a technique where many flares are superposed in order to get a mean profile showing all typical stable features. The authors applied this technique to \ginga\ data of the XRB Cyg\,X-1 and found multicomponent characters of the averaged shot. The latter has a central spike with two humps on both sides of the central spike.

The averaged shot profile method suggests that individual flares are superposed on each other. Such a process called shot noise has additive character and do not produce the observed linear rms-flux relation. However, as mentioned above, the flickering shows this linearity which is typical for a multiplicative process. This has strong theoretical consequences (see e.g. \citealt{uttley2005}). Therefore, the averaged shot profile method together with shot noise model can be used just for purely phenomenological purposes (see e.g. \citealt{bruch2015}).

High cadence, long and continuous light curves from the \kepler\ satellite (\citealt{borucki2010}) are an excellent opportunity for such shot profile study, because the data offer hundreds of individual flares. Such a systematic study using these data with unprecedented quality was made by \citet{sasada2017} in the case of the blazar W2R\,1926+42. The authors averaged 195 individual flares and made several test to prove the reality of the detected features. The superposed shot profile consists of three components, i.e. a central spike and two side-lobes on each side of the spike.

Another accreting system studied in details thanks to \kepler\ data is the CV MV\,Lyr. Its PDS has four components (\citealt{scaringi2012a}) with the highest one probably generated by the inner evaporated hot geometrically thick and optically thin corona above a geometrically thin optically thick accretion disc (\citealt{scaringi2014a}), the so-called sandwich model. If the variability is generated by the corona radiating hard X-rays, the optical \kepler\ data is a result of the reprocessing of the X-rays, and the corresponding PDS component must be detected in X-rays too. This interpretation was confirmed by \xmm\ observations (\citealt{dobrotka2017}) where two highest PDSs components were detected.

The physical model well fitting both the PDS features and the linear rms-flux relation of MV\,Lyr (\citealt{scaringi2014a}) is the accretion fluctuation propagation scenario (\citealt{lyubarskii1997}, \citealt{kotov2001}, \citealt{arevalo2006}). Following this model every accretion rate fluctuation generated anywhere in the disc is propagating inside. Further fluctuations are generated during this travel, and all these fluctuations "summed" during the way modulate the inner mass accretion rate.

Due to the complex multifrequency study of the system MV\,Lyr performed until now by various instruments and authors, the CV is an ideal target for the mentioned shot profile study in order to get additional information. The main motivation of such study is the comparison with the AGN W2R\,1926+42 and XRB Cyg\,X-1. In this paper we perform this study and try to compare all three very different objects in nature, but all having accretion process as the main engine powering their radiations.

\section{Data}

\begin{figure}
\resizebox{\hsize}{!}{\includegraphics[angle=-90]{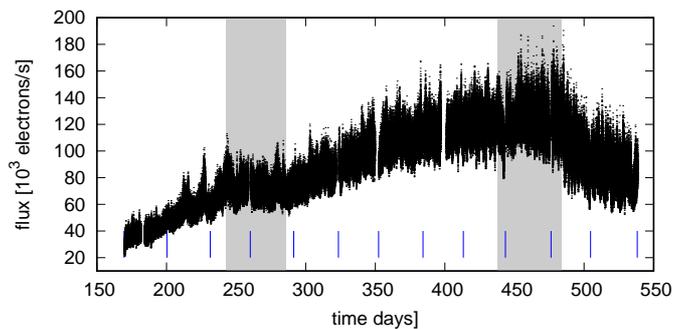}}
\caption{Analysed \kepler\ light curve of MV\,Lyr with shaded regions roughly showing two intervals with constant flux. The vertical blue lines divide the light curve into 12 subsamples used for shot profile evolution calculation in Section~\ref{section_evolution}.}
\label{lc}
\end{figure}

For our study we selected a part of the data already presented and studied by \citet{scaringi2012a}. The data are taken by the \kepler\ satellite (\citealt{borucki2010}) with a cadence of approximately 60\,s. Our light curve lasts approximately 370 days and comprises more or less monotonically increasing and subsequently decreasing trends. Fig.~\ref{lc} depicts the light curve with two shaded regions roughly showing two intervals with constant flux.

\section{Superposed shot profile}
\label{section_shot_profile}

We performed a similar superposed shot profile analysis as presented by \citet{negoro1994}. Our work is motivated by the use of this technique to \kepler\ data of the AGN W2R\,1926+42 performed by \citet{sasada2017}. Our technique is slightly different, and has two steps. First is the peak identification, and second the flare extension selection. For the former we used a simple condition, that a light curve point is identified as a peak, if $N_{\rm pts}$ points to the left and $N_{\rm pts}$ points to the right have lower fluxes than the tested point. The second step is the flare extension selection in order to not superimpose a declining branch of one flare with a rising branch of the adjacent flare, and vice versa. For this purpose a flare is identified as $N_{\rm ptsext} = N_{\rm pts}/2$ points to the left and $N_{\rm ptsext} = N_{\rm pts}/2$ to the right from the peak point. We performed some further tests which we present bellow, but this algorithm is the most secure in order to average individual flares.

After the flare selection we performed a simple averaging of the flare points, and resulting averaged flux minimum was subtracted from all averaged points. All flares with rare individual null points were excluded from the averaging process. The best would be to choose a short interval of the light curve with more or less constant flux (for example the first shaded area in Fig.~\ref{lc}), but this would result in a too low flare number. Therefore, we first used the rising part of the light curve from the beginning till day 439 (until the beginning of the second shaded area in Fig.~\ref{lc}). A fainter division of the light curve yields a lower flare number, but is suitable for an evolution study.

Finally, the long-term trend visible in Fig.~\ref{lc} has no effect on the results. De-trended data yields the same profile, because any long-term trend is negligible within the time extent of a single shot. Moreover, \kepler\ data are not uniformly spaced because of barycentric correction. Our averaging method assumes evenly spaced data, and every shot is handled separately. Since the central points (peaks) are aligned, any time step modification has effect within half of a single shot. The largest time extent of a single shot used in this work is of 3.3\,h, which is extremely short for a time step variation due to barycentric correction. A simple test with evenly re-sampled data using linear interpolation yields almost the same shot profile with negligible differences. However, using linear interpolation we always get lower variance (the interpolated point is always between the two real points, both the time and flux values). The latter is the reason why we prefer original data.

\begin{figure}
\resizebox{\hsize}{!}{\includegraphics[angle=-90]{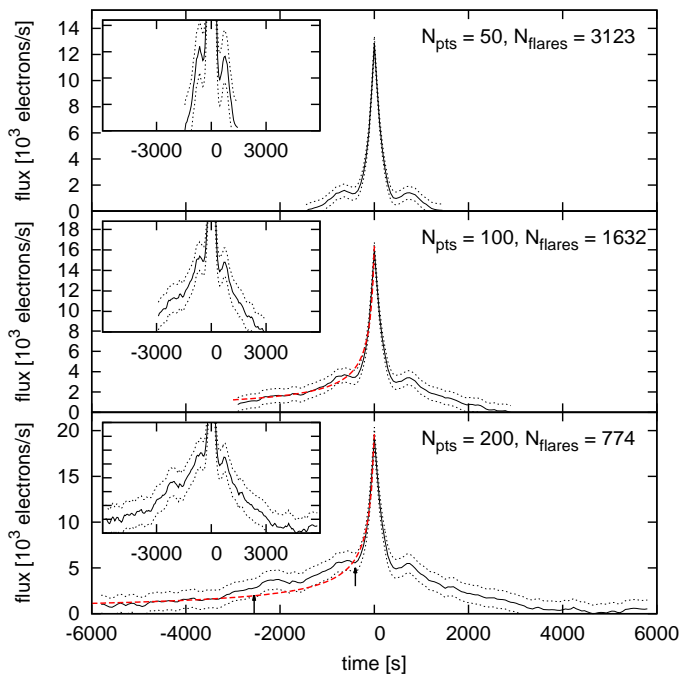}}
\caption{Superposed shot profiles using three different values of $N_{\rm pts}$ resulting in different flare numbers $N_{\rm flares}$ and flare time extension. The inset panels are detailed views without the central spike in order to better visualize the side-lobes. Solid black line represents the mean value, while the dotted thin line represents the standard error of the mean. The thick dashed (red) line is the power law fit and the two arrows in the bottom panel show the interval excluded from the fitting process (see Section~\ref{discussion} for details).}
\label{profile_mean}
\end{figure}

Three examples of the superposed averaged shot profile are shown on Fig.~\ref{profile_mean} with inset panels as a zoom. The shot profile consists of a central spike and apparent side-lobes on both sides. Another structure is visible at approximately -2100\,s (in the middle and bottom panel). The best would be to have a $N_{\rm ptsext}$ parameter as large as possible to get large time extension before and after the central spike to see all details. But, as seen in Fig.~\ref{profile_mean}, a higher $N_{\rm pts}$ with larger time extension yields lower flare number resulting in a more noisy shot profile. Therefore, some compromise is needed between the flare profile time extension and data scatter.

\begin{figure}
\resizebox{\hsize}{!}{\includegraphics[angle=-90]{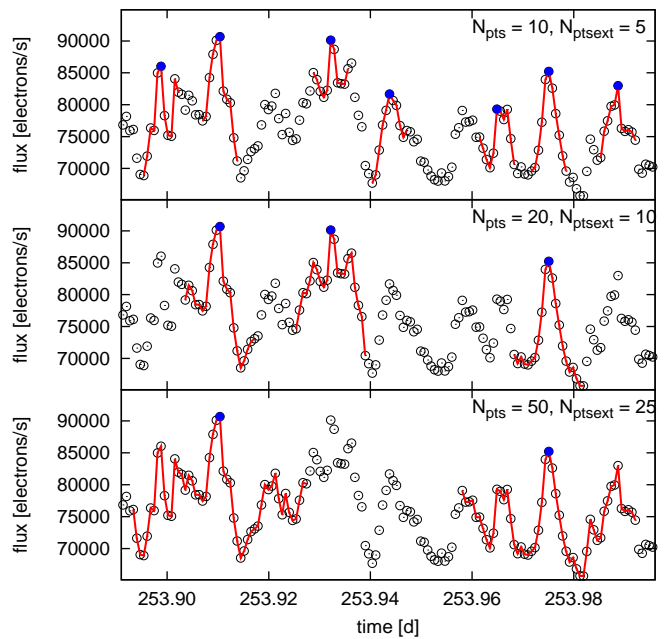}}
\caption{Example of selected flares (red solid lines) using different $N_{\rm pts}$ values. The open circles represent the light curve data, while the solid (blue) circles represent the maxima.}
\label{lc_flares}
\end{figure}

The presented technique is different from the one used by \citet{negoro1994} and \citet{sasada2017}. These authors superposed well resolved and "isolated" flares (see Fig.~4 in \citealt{sasada2017}). However, in the MV\,Lyr \kepler\ data such well resolved and "isolated" flares are not present, but many flares are superposed instead. This is depicted in Fig.~\ref{lc_flares}. Different $N_{\rm pts}$ were used to select the flares. In the case of $N_{\rm pts} = 10$ the majority of flares are well resolved and have a spiky shape. However, such $N_{\rm pts}$ value is too low to study any expanded structures because of short time extension. When increasing the $N_{\rm pts}$ parameter, the resolved flares become too complicated and many superposed flare maxima can be present in the selected data interval. The averaging process keeps the central spike, and smooths out all randomly present adjacent flares maxima and keep only the real structures. This is also important in uncertainty determination. We used the standard error of the mean instead of the standard deviation, because the latter would describe the data scatter due to superposed adjacent flares maxima, and not the intrinsic profile uncertainty.

\subsection{Reality test}
\label{reality_test}

In order to exclude any doubts due to possible numerical artifacts, we performed a simple reality test to prove that the detected substructures are real. We took the largest shot profile ($N_{\rm pts} = 200$) from Fig.~\ref{profile_mean} and constructed a synthetic light curve with the same duration and sampling as the observed one. 100000 flares were superposed at random in order to construct the light curve. The resulting artificial flux was rescaled in order to be comparable with observed light curve characteristics (mean flux and rms). Such a process is not ideal because the superposition of flares is a shot noise model not satisfying all typical features of the real light curves, but our goal is not reproduction of real data, but testing whether the superposition of many flares keeps the original shot profile. It shows that every structure present in the input shots is present in the resulting averaged profile with secure $N_{\rm pts}$ selection. This is an important test because of the light curve character, where many adjacent flares maxima are superposed in the selected flare region. The presence of such adjacent maxima do not influence the result then.

\begin{figure}
\resizebox{\hsize}{!}{\includegraphics[angle=-90]{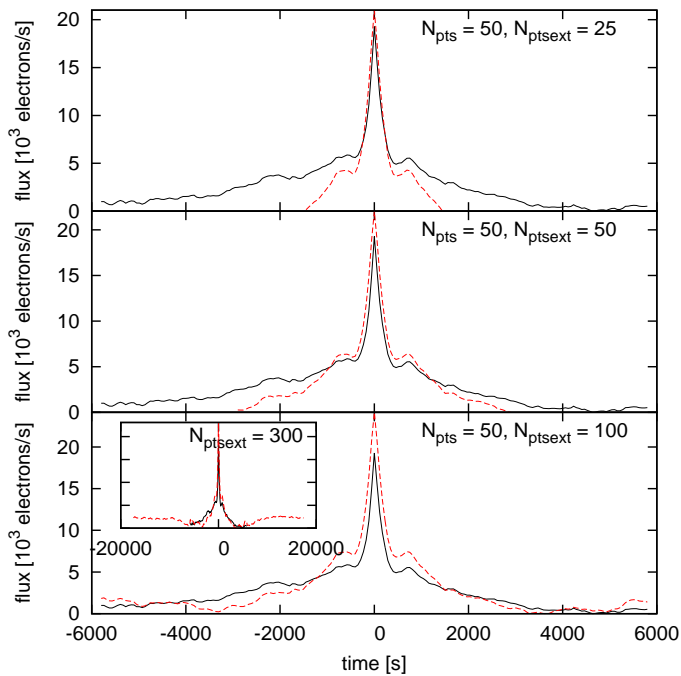}}
\caption{Superposed shot profiles from simulated light curves using different extraction parameters $N_{\rm ptsext}$ (dashed red line). As input shot profile (black solid line) for synthetic light curve construction we used the one from bottom panel of Fig.~\ref{profile_mean}.}
\label{profile_simul}
\end{figure}

The test results are shown as red lines in Fig.~\ref{profile_simul}. We used a selection criterion of $N_{\rm pts} = 50$, but we used different $N_{\rm ptsext}$ values to investigate the flare extension. The averaged shot profile has the same structure, i.e. a central spike with side-lobes, and a possible hump at -2100\,s for $N_{\rm ptsext} \geq 50$. However, noticeable is the increasing/decreasing trend of the flare wings. Its behaviour depends on the parameter $N_{\rm ptsext}$, i.e. whether it is lower or equals to $N_{\rm pts}$. With $N_{\rm ptsext} = 2 \times N_{\rm pts}$ the flare wings start to change trend at approximately -3500 and 5000\,s, they rise instead of decline and vice versa. For an extreme value of $N_{\rm ptsext} = 6 \times N_{\rm pts}$ this false trend is clearer which stabilizes itself above 10000\,s.

This test suggests that $N_{\rm ptsext}$ can be set up as equal to $N_{\rm pts}$, but in order to avoid any non detected artifacts resulting from superimposition (repetition of the same data) of adjacent flares we used $N_{\rm ptsext} = N_{\rm pts}/2$ as already mentioned. The large quantity of \kepler\ data allows such waste.

Finally, when using a much simpler profile for input shots for the synthetic light curve construction (a simple spike without side-lobes), only a profile similar to the used one was obtained as the averaged profile. Therefore, the side-lobes are real and are not an artifacts.

\section{Shot profile fitting}

In order to quantitatively describe the detected profile we concentrated our study to the central spike and to the most dominant side-lobes. These features are well resolved even in shorter light curve subsegments where lower $N_{\rm pts}$ is required in order to get larger flares quantity. We fitted these two features individually using {\small GNUPLOT}\footnote{http://www.gnuplot.info/} software yielding the fitted parameters with the standard errors.

\subsection{Central spike}
\label{central_spike}

\begin{figure}
\resizebox{\hsize}{!}{\includegraphics[angle=-90]{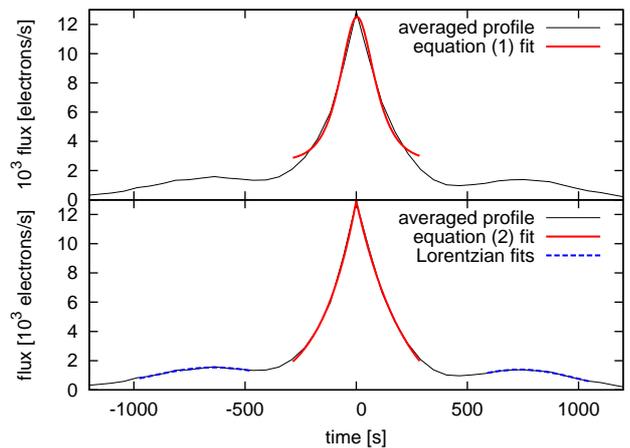}}
\caption{Fits of the individual shot profile components. Black line is the $N_{\rm pts} = 50$ case with best resolution from upper panel of Fig.~\ref{profile_mean}. Red line is the exponential fit following equation (\ref{wrong_fit}) or \ref{correct_fit}), and blue lines are the Lorentzian fits using equation (\ref{sidelobe_fit}).}
\label{profile_fit}
\end{figure}

\citet{sasada2017} fitted the central spike with two functions. The first was proposed by \citet{abdo2010} to represent blazar flare profiles, and has a form of
\begin{equation}
F(t) = F_{\rm c} + F_{\rm 0} \left(e^{-t/T_{\rm r}} + e^{t/T_{\rm d}} \right)^{-1},
\label{wrong_fit}
\end{equation}
where $t$ is time, $T_{\rm r}$ and $T_{\rm d}$ are variation times scales of a rising and declining branch, respectively. $F_{\rm c}$ and $F_{\rm 0}$ represents the constant level and the amplitude of the shots, respectively. Second fit used by \citet{sasada2017} has a form of
\begin{equation}
F(t) = \left\{
\begin{array}{ll}
F_{\rm c} + F_{\rm 0} e^{t/T_{\rm r}}, & t < 0\\
F_{\rm c} + F_{\rm 0} e^{-t/T_{\rm d}}, & t > 0,
\label{correct_fit}
\end{array}
\right.
\end{equation}
with the same parameter meanings as in Eq.~(\ref{wrong_fit}). Fig.~\ref{profile_fit} displays both fits as red lines performed from\footnote{The limits are chosen empirically.} -285 to 285\,s. Clearly, the case described by Eq.~(\ref{correct_fit}) is better\footnote{The sum of residuals square is 6 times lower than with the Equation~(\ref{wrong_fit}).} and even describing the profile well. The most expressive difference is visible near the central spike maximum, where the model following equation (\ref{wrong_fit}) do not describe the pointed profile well. The fitted time scales with the ratios $T_{\rm r}/T_{\rm d}$ are listed in Table~\ref{fit_param}.
\begin{table}
\caption{Fitted time scales of rising and declining parts of the central spike, and the corresponding time scale ratios.}
\begin{center}
\begin{tabular}{lcccc}
\hline
\hline
data & $T_{\rm r}$ & $T_{\rm d}$ & $T_{\rm r}/T_{\rm d}$\\
& (s) & (s) &\\
\hline
all & $164.1 \pm 19.9$ & $225.6 \pm 34.5$ & $0.73 \pm 0.14$\\
subsample 1 & $165.1 \pm 34.1$ & $205.3 \pm 43.6$ & $0.80 \pm 0.24$\\
subsample 2 & $171.8 \pm 19.7$ & $211.6 \pm 39.4$ & $0.81 \pm 0.18$\\
subsample 3 & $186.0 \pm 30.5$ & $317.2 \pm 48.5$ & $0.59 \pm 0.13$\\
subsample 4 & $154.2 \pm 23.7$ & $324.8 \pm 45.0$ & $0.47 \pm 0.10$\\
subsample 5 & $141.4 \pm 18.1$ & $236.3 \pm 57.6$ & $0.60 \pm 0.16$\\
subsample 6 & $157.3 \pm 08.1$ & $188.5 \pm 29.4$ & $0.83 \pm 0.14$\\
subsample 7 & $157.0 \pm 17.9$ & $233.5 \pm 33.7$ & $0.67 \pm 0.12$\\
subsample 8 & $227.1 \pm 83.4$ & $214.6 \pm 10.6$ & $1.06 \pm 0.39$\\
subsample 9 & $196.0 \pm 38.1$ & $168.2 \pm 27.6$ & $1.17 \pm 0.30$\\
subsample 10 & $150.7 \pm 46.3$ & $180.7 \pm 27.9$ & $0.83 \pm 0.29$\\
subsample 11 & $148.0 \pm 63.6$ & $184.4 \pm 42.1$ & $0.80 \pm 0.39$\\
subsample 12 & $160.3 \pm 50.7$ & $199.6 \pm 28.9$ & $0.80 \pm 0.28$\\
\hline
\end{tabular}
\end{center}
\label{fit_param}
\end{table}

\subsection{Side-lobes}
\label{sidelobes}

We further investigated the time "location" of the side-lobes. For this purpose we fitted them with a Lorentz function
\begin{equation}
\Psi = a + L(b,\Delta,t,t_{\rm 0}),
\label{sidelobe_fit}
\end{equation}
\begin{equation}
L(b,\Delta,t,t_{\rm 0}) = \frac{b \Delta}{\pi} \frac{1}{\Delta^2 + (t - t_{\rm 0})^2},
\label{lorentzian}
\end{equation}
($\Psi$ is the flux, $a$ and $b$ are constants, $\Delta$ is the half-width at half maximum, $t$ is time from the peak and $t_{\rm 0}$ is the searched side-lobe time location) to the visually selected lobes yielding the times of $-639.1 \pm 13.1$ and $736.2 \pm 6.4$\,s for the rising and declining lobe, respectively. The time distance between the lobes is approximately 1375\,s and the fits are shown as blue lines in Fig.~\ref{profile_fit}.

\subsection{Components evolution}
\label{section_evolution}

\begin{figure}
\resizebox{\hsize}{!}{\includegraphics[angle=-90]{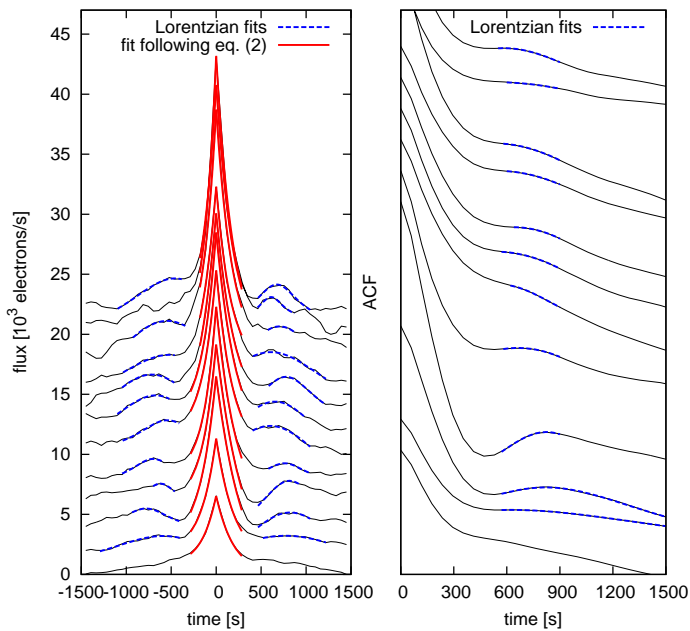}}
\caption{Left panel - Individual shot profiles ($N_{\rm pts} = 50$) from light curve subsamples with individual component fits. The time evolution is from bottom to top. The lowest profile has the original flux values, while every next profile is offset by 2000 el./s vertically. Right panel - The same as left panel but derived using ACF.}
\label{profile_evol}
\end{figure}

\begin{figure}
\resizebox{\hsize}{!}{\includegraphics[angle=-90]{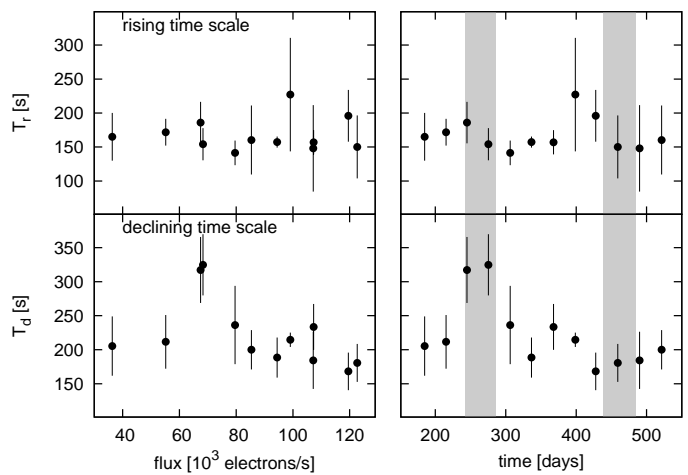}}
\caption{Time scales of the central spike evolution as function of averaged flux (left panels) and of light curve segment time (right panels). The shaded areas in the right panels correspond to the shaded areas in Fig.~\ref{lc}.}
\label{frekv_evol_spike}
\end{figure}

\begin{figure}
\resizebox{\hsize}{!}{\includegraphics[angle=-90]{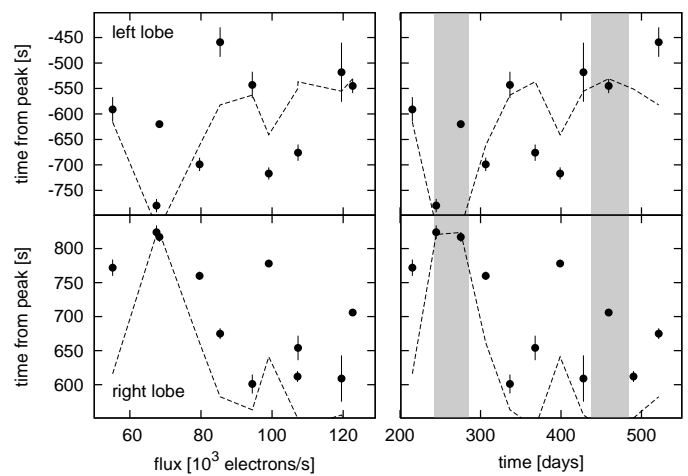}}
\caption{The same as Fig.~\ref{frekv_evol_spike}, but for side-lobes central times. The dashed lines represent the same, but derived from the autocorrelation functions in Section~\ref{acf} for comparison.}
\label{frekv_evol_lobes}
\end{figure}

For the profile evolution we used light curve subsamples as they come from the \kepler\ archive (the intervals are marked as blue lines in Fig.~\ref{lc}), and the fitted profiles are shown in the left panel of Fig.~\ref{profile_evol}. The evolution of fitted parameters as time scales and side-lobes peak times are depicted in Figs.~\ref{frekv_evol_spike} and \ref{frekv_evol_lobes}. We used two regimes; 1) fitted parameter vs mean flux of the data subsample, and 2) fitted parameters vs central time of the light curve subsample.

For the spike fitting we changed the fitted time intervals when the fits run away significantly from the shot profile, otherwise we used the same time interval as in Section~\ref{central_spike}. This change of the fitting interval was needed in four last cases with largest fluxes, because of too high amplitude of the central spike. The fitted time scales with the ratios for individual light curve subsamples are listed in Table~\ref{fit_param}.

Inspecting the Fig.~\ref{frekv_evol_spike} we see nothing significant when studying the flux or time evolution, except two points in both bottom panels. The declining time scale $T_{\rm d}$ shows a significant value deviation for fluxes of approximately 70 el./s. This value corresponds to the local flux plateau shown as shaded area in Fig~\ref{lc}. The second flux plateau at the maximum flux of the light curve, where the trend changes from rising to declining, does not show this $T_{\rm d}$ deviation.

Not every averaged shot profile has discernible side-lobes. The first profile with the lowest flux has missing side-lobes, while the rising lobe of the 11$^{th}$ case is too noisy for any relevant fit. The resulting side-lobes central times in Fig.~\ref{frekv_evol_lobes} are too scattered to draw any conclusion from the flux or time evolution.

\section{Autocorrelation}
\label{acf}

Another way how to study the fast variability shape or the shot profile is the use of the autocorrelation function (ACF). However, the ACF is an even/symmetric function not suitable to study asymmetric profiles, because the location of the two side-lobes is represented by only one feature in the ACF. Anyhow, it is worth investigating this way in order to get another reality test of the shot profile method and derived shot substructures.

In order to calculate the ACF we need evenly spaced data which is a problem using whole \kepler\ light curve. Every rare null point or larger gap we replaced by linear interpolation. An ACF of the data used in Fig.~\ref{profile_fit} is depicted in Fig.~\ref{acf}. There is a significant lobe located at $657.25 \pm 4.07$\,s shown as the Lorentz fits as in Section~\ref{sidelobes}. Apparently, this value is very close to the average (687.7\,s) of both side-lobe locations derived by the shot profile method. This suggests the reality of the substructures. Furthermore, there is a weak but noticeable hump at $2200 \pm 4$\,s in the ACF (enhanced by linear detrending in the inset panel of Fig.~\ref{acf}). This feature\footnote{This feature is of very low significance, but we do not use it for any analysis. Therefore, we do not investigate or discuss its credibility.} location is similar to the left hump in the shot profile in bottom panel of Fig.~\ref{profile_mean}.
\begin{figure}
\resizebox{\hsize}{!}{\includegraphics[angle=-90]{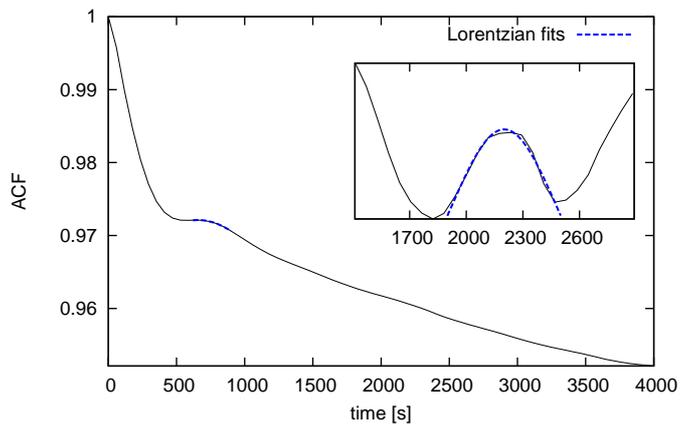}}
\caption{The same as in Fig.~\ref{profile_fit}, but derived using ACF and for larger time scales (lags). The inset panel shows ACF region between 1400 and 2900\,s after linear detrending in order to enhance the hump structure fitted by Lorentzian.}
\label{acf}
\end{figure}

The right panel of Fig.~\ref{profile_evol} shows the ACF evolution as an analogy to the left panel. Worth noting is the first (bottom) ACF not showing a side-lobe, and the fourth ACF with a clearly enhanced side-lobe. Both results are consistent with the shot profile evolution. Fig.~\ref{frekv_evol_lobes} compares the ACF lobe location evolution with the ones derived from the averaged shot profile method. First view does not suggest a perfect or strong correlation, but Kendall rank correlation coefficients for the rising and declining side-lobes are 0.87 and 0.69, respectively. This suggests significant correlation (even strong correlation in the rising lobe case) of the results derived from both methods.
%
%

The non-even character of the \kepler\ data must be tested also in the case of the ACF. Using linearly interpolated data like in the shot profile calculation, we get the same ACF features, just the values are slightly vertically offset toward higher values. Therefore, we used original data like in Section~\ref{section_shot_profile}.

Finally, the presence of the side-lobes, showing the similar behaviour/evolution compared to the shot profile method, is an additional proof that the detected features are real, and the averaged shot profile method is working well.

\section{Power density spectra}
\label{pds_section}

\subsection{Simulated PDSs}
\label{pds_section_simul}

In Section~\ref{sidelobes} we showed that the two side-lobes are located at times $-639.1 \pm 13.1$ and $736.2 \pm 6.4$\,s. These side-lobes correspond to two waves of a $687.7 \pm 7.3$\,s-period (average value) with a corresponding frequency of log($f$/Hz) = $-2.837\pm 0.004$. The latter suggests a similarity with frequency components detected in PDSs studied by \citet{scaringi2012a}. In order to study the meaning of this similarity, we simulated light curves with the simple shot noise process\footnote{The same procedure as in Section~\ref{reality_test}.} using the shot profile from Fig.~\ref{profile_mean} (we used the one with the largest time scale, bottom panel). We performed PDS analysis of this synthetic light curves in the same way as in \citet{dobrotka2015} to get closest approach to the PDSs presented in \citet{scaringi2012a}. We selected light curve samples with duration of 25\,d, we divided these samples into 5 equal subsamples, for every subsample we calculated a periodogram using the Lomb-Scargle (\citealt{scargle1982}) algorithm, transformed the periodogram into log-log scale, averaged all 5 log-log\footnote{Following \citet{papadakis1993} it is more suitable to average log($p$) instead of power $p$.} periodograms and binned the averaged periodogram into equally spaced bins in order to get the PDS estimate. The Fourier transform would be ideal for a perfectly equidistant simulated light curve, but for the later purpose where we calculate the PDSs from the \kepler\ data we choose Lomb-Scargle as ideal for non equidistant data. \kepler\ data have a lot of gaps and spurious null points or intervals of null points. Therefore, we use the Lomb-Scargle method for all PDS estimates to get equivalent results. Finally, Lomb-Scargle is often used for PDS study also in equidistant data (see ex. \citealt{shahbaz2005}).

\begin{figure}
\resizebox{\hsize}{!}{\includegraphics[angle=-90]{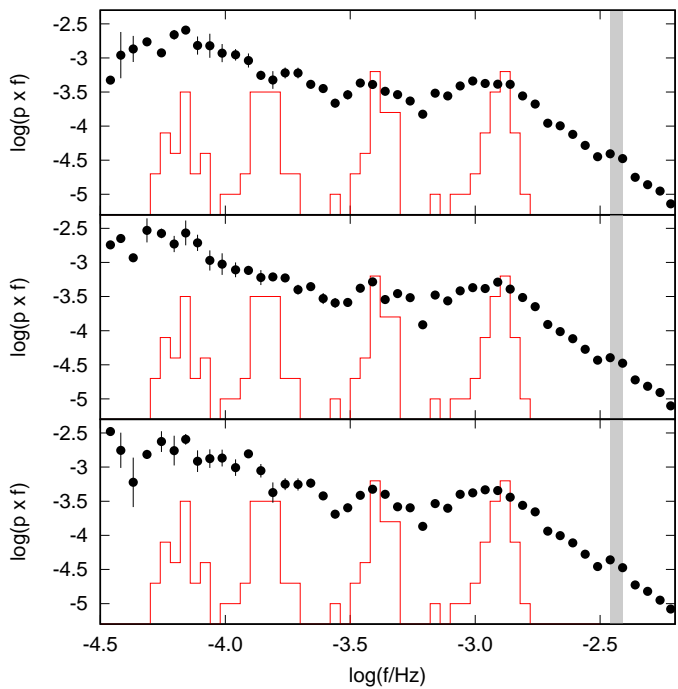}}
\caption{PDSs calculated from simulated light curve subsamples. The points are averaged means with the standard error of the mean as vertical lines. The vertical shaded area shows the frequencies of two points with a small power deviation at log($f$/Hz) = -2.46 and -2.41 representing a potential high frequency PDS component. The solid red lines represent the histogram of frequencies detected in \kepler\ data from Table~1 of \citet{scaringi2012a}.}
\label{pds_simul}
\end{figure}

Three examples of simulated PDSs are shown in Fig.~\ref{pds_simul}. The clearest pattern in all the three cases is the break frequency or PDS component at approximately log($f_1$/Hz) = -3. Another common well resolved "hump" is seen at around log($f_2$/Hz) = -3.4. Lower frequencies are rather flat and scattered without any dominant pattern. Some power excess is noticeable at approximately log($f_3$/Hz) = -3.7, and the lower end of the PDS shows common power decrease toward low frequencies from approximately log($f_4$/Hz) = -4.2. But these are of low significance. We compared these simulated PDSs with frequencies\footnote{$f = (\nu_0^2 + \Delta^2)^{1/2}$} detected in \kepler\ data from Table~1 (see also Fig.~4) of \citet{scaringi2012a}. The two highest frequencies from \kepler\ data clusters around the clearest PDS patterns at $f_1$ and $f_2$, and a possible match is noticeable in the $f_4$ case. There is no obvious or systematic correlation of $f_3$ with the observed histogram, even the presence of any power excess close to $f_3$ in the simulations is not certain (middle panel). Therefore, we can conclude that at least the two highest PDS components detected by \citet{scaringi2012a} are directly "seen" in the averaged shot profile.

Moreover, two deviated points are seen at log($f$/Hz) = -2.46 and -2.41 in the simulated PDSs (marked as the vertical shaded area in Fig.~\ref{pds_simul}). If real, these suggest a new high frequency PDS component not detected so far with a characteristic frequency in between the two values.

\subsection{\kepler\ PDSs}

\begin{figure}
\resizebox{\hsize}{!}{\includegraphics[angle=-90]{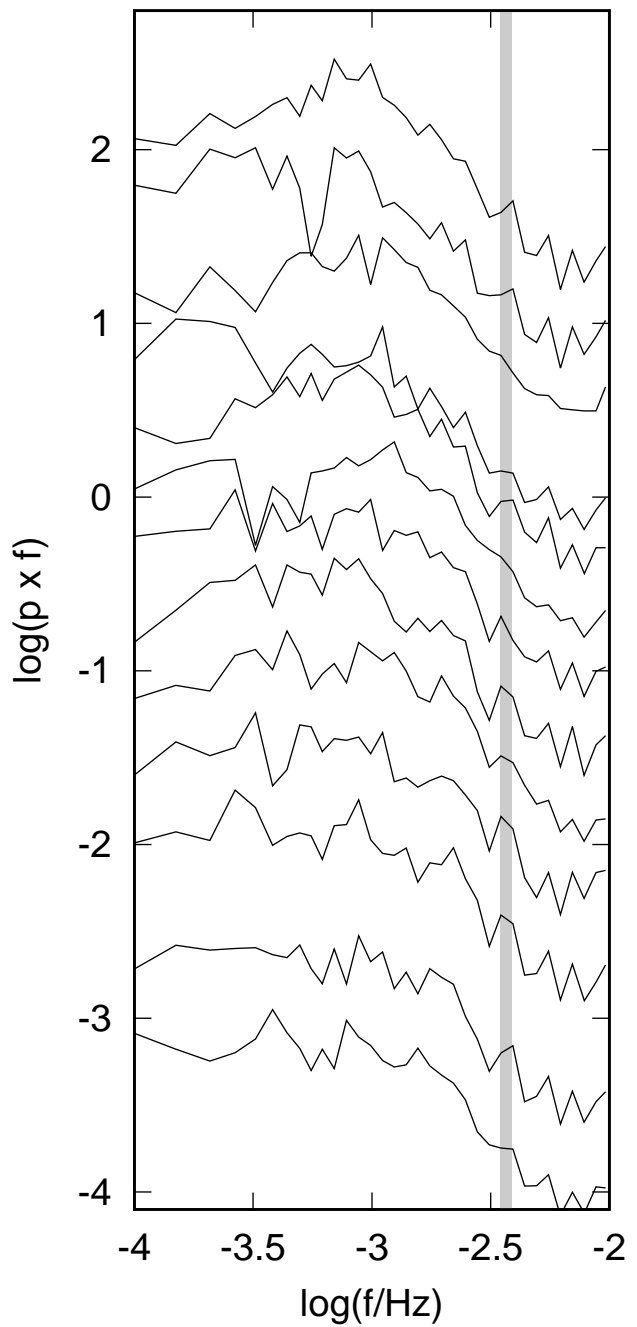}}
\caption{PDSs calculated from \kepler\ light curve. The whole light curve in Fig.~\ref{lc} is divided into equal 25 day segments and every PDS represents the corresponding segment (the first PDS is at the bottom). The vertical shaded area shows the frequencies of expected power excess between log($f$/Hz) = -2.46 and -2.41.}
\label{pds_kepler}
\end{figure}

The two deviated points with very stable values motivated us to search for this feature in the real data. We first divided the whole studied \kepler\ light curve (Fig.~\ref{lc}) into 25 day subsamples and performed the same Lomb-Scargle analysis as in Section~\ref{pds_section_simul}. Fig.~\ref{pds_kepler} shows these PDSs with vertical dotted lines showing the two frequencies.

Not every PDS from the observed data shows a power deviation at the frequency of the expected power excess, but some cases are promising. However, the significance is very low which is expected also from simulations. The Poissonian noise in the real data makes the detection even harder.

\subsection{\xmm\ PDS}
\label{pds_section_xmm}

\begin{figure}
\resizebox{\hsize}{!}{\includegraphics[angle=-90]{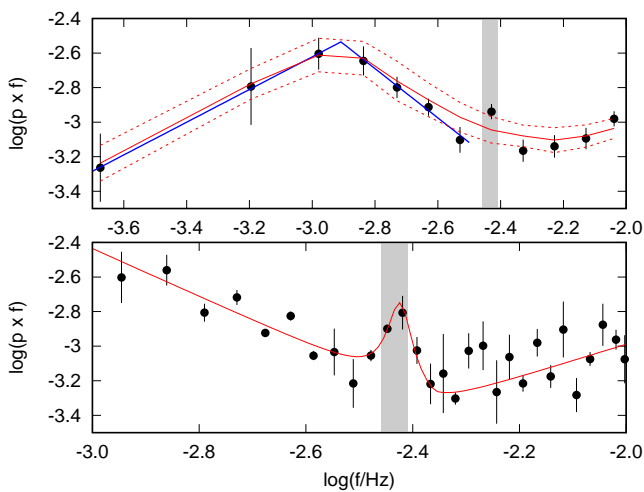}}
\caption{Upper panel - PDS of \xmm\ data with broken power law fit, and mean (solid red line) PDS value with 1-$\sigma$ intervals (red dashed lines) from one hundred simulations. Lower panel - PDS with higher resolution and broken power law fit plus Lorentzian. The shaded area shows the region of expected power excess between log($f$/Hz) = -2.46 and -2.41.}
\label{pds_xmm}
\end{figure}

As a next step we searched for the same component in \xmm\ data. We took the PN light curve from \citet{dobrotka2017}. We performed the same lomb-Scargle calculation as in the previous cases, but we rebined the averaged periodogram into larger bins (interval of 0.1 in log scale), and with a minimum number of averaged points of 5 (upper panel of Fig.~\ref{pds_xmm}). We did so to smooth strongly the periodogram in order to see the main trends and the strongest features. A considerably deviated point at approximately log($f$/Hz) = -2.43 (exactly between the two frequencies from simulated PDSs) is seen.

As a reality test, we simulated light curves using the method of \citet{timmer1995}. This method uses an input PDS to generate noisy light curves. For this purpose we fitted the low resolution PDS with a broken power law\footnote{A power law has a form of log($p$) = $a$\,log($f$) + $b$, where $a$ is the power low slope and $b$ is the constant. Broken power law means, that two different power laws are used bellow and above a break frequency.} (blue thick solid line) to get PDS estimate describing the main trends. The simulated light curves have the same length, sampling, mean flux value and variance as the observed data. We added a Gaussian noise to the curve to fit well the highest frequency part of the observed PDS. The solid red line in the upper panel of Fig.~\ref{pds_xmm} is the mean value of 100 simulated light curves, and the dashed curves in represent the 1-$\sigma$ interval. Apparently, the anomalous point deviates from the main trend with amplitude slightly larger than the 1-$\sigma$ interval.

This power deviation can represent a new high frequency PDS component. In order to search for its characteristics, we increased the PDS resolution from 0.1 to 0.025 with a minimum number of points 3. We fitted the high resolution PDS with a broken power law plus a Lorentzian (Eq.~\ref{lorentzian}, with $L = {\rm log}(p)$, $t = f$ and $t_0 = f_0$). The fit is depicted in the bottom panel of Fig.~\ref{pds_xmm} and the fitted characteristic frequency is log($f_0$/Hz) = $-2.424^{+0.007}_{-0.008}$.

\section{Discussion}
\label{discussion}

We performed analysis of the superposed shot profile of the \kepler\ data of the nova like system MV\,Lyr based on the original idea of \citet{negoro1994}. We investigated the details of the shot profile and its time evolution and we searched for links between the shot profile and the multicomponent PDS of the \kepler\ data.

\subsection{Shot profiles from \kepler\ data}
\label{shot_kepler}

This work is motivated by a similar analysis of \kepler\ data of the blazar W2R\,1926+42 performed by \citet{sasada2017}. The authors found very similar shot profile as depicted in our Fig.~\ref{profile_mean}, i.e. an asymmetric central spike with two side-lobes. The fitting of the central spike in MV\,Lyr yields a time scale ratio $T_{\rm r}/T_{\rm d}$ of $0.73 \pm 0.14$ which agrees well with the ratio of $0.70 \pm 0.03$ for the blazar W2R\,1926+42 (from Table~1 of \citealt{sasada2017}). Furthermore, \citet{sasada2017} subdivided the light curve into subsamples yielding time scale ratios of 0.40 - 1.01. This compares to 0.47 - 1.17 in the case of MV\,Lyr (Table~\ref{fit_param}). The weighted means for the former and latter is $0.63 \pm 0.11$ and $0.84 \pm 0.13$, respectively. Apparently, the asymmetry of the central spike characterized by these ratios is very similar for both objects. The spike has fast rise and slow decline. Similar non-symmetry in MV\,Lyr was suggested by \citet{scaringi2014b}. However, the authors report the opposite, i.e. the flare is rising more slowly than it is falling. This asymmetry is derived from the Fourier analysis, and it is valid for the highest break frequency at log$(f/{\rm Hz}) = -3$ in the PDS. This discrepancy requires detailed comparison between the PDS and the shot features (see bellow).

The timing characteristics of the central spike can be estimated from the two time scales $T_{\rm r}$ and $T_{\rm d}$ (from Table~\ref{fit_param}). Following \citet{negoro2001} these time scales correspond to knees in the PDS with corresponding frequencies of $1/(2 \pi T)$. \citet{sasada2017} found a break frequency of log($f$/Hz) = -4.39 in the PDS of W2R\,1926+42 which is very close to the rising and declining time scales of the central spike of the blazar data. Therefore, this frequency represents the central spike of the mean profile. Following \citet{scaringi2012a} the observed PDS of MV\,Lyr has the highest frequency component of approximately log($f$/Hz) = -3. Based on the previous analogy, such frequency should be seen in the shot profile. Calculating the characteristic frequencies using $T_{\rm r}$ and $T_{\rm d}$ we get values of log($f$/Hz) = -3.01 and -3.15, respectively. Apparently, the dominant central spike of the averaged shot profile corresponds to the dominant PDS component in MV\,Lyr.

Furthermore, in Section~\ref{pds_section} we show that a pattern with a frequency of log($f$/Hz) = -2.42 is generated in the simulated PDS (Fig.~\ref{pds_simul}). This motivated us to search for this feature in the real data. While the \kepler\ PDSs show the expected power excess with very low significance, the excess in the re-analysed \xmm\ PDS is more convincing. A corresponding time scale is of 263\,s. Such a structure is not identifiable in our shot profile, but such a short time scale can corresponds to the narrowest peak of the central spike.

\begin{figure}
\resizebox{\hsize}{!}{\includegraphics[angle=-90]{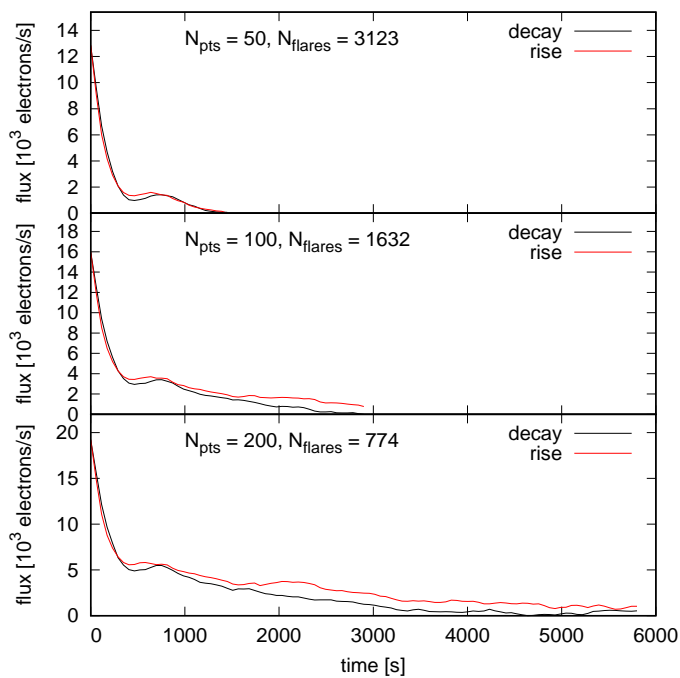}}
\caption{The same as Fig.~\ref{profile_mean}, but with directly compared rise and decay parts of the mean profiles.}
\label{profile_mean_rise-dec_compar}
\end{figure}

Moreover, the side-lobes maxima in MV\,Lyr are located at times $-639.1 \pm 13.1$ and $736.2 \pm 6.4$\,s from the zero. As already mentioned these side-lobes correspond to two wave-lengths of a $687.7 \pm 7.3$\,s-length wave with corresponding frequency of log($f$/Hz) = $-2.837\pm 0.004$. The latter is very close to the dominant PDS feature at log($f$/Hz) = -3, and clearly it contributes to it. Apparently all the fine structures of the shot profile have corresponding components in the PDS. However, while the individual components are blending and are seen as just one dominant PDS feature or are non-distinguishable due to low resolution, the shot profile is able to recognize them separately. This can explain the discrepancy between asymmetry of the dominant central spike detected in this work and opposite asymmetry derived by \citet{scaringi2014b}, i.e. the latter reported positive skewness for a dominant PDS break frequency at approximately log($f$/Hz) = -2.7. This is close to the side-lobe frequency of log($f$/Hz) = $-2.8$, while for the lower frequencies of log($f$/Hz) = -3.01 and -3.15 related to the central spike the time skewness is negative (see Fig.~1 of \citealt{scaringi2014b}). For even lower frequencies the skewness returns back to positive values. Such negative skewness agrees with the slower decay as seen in the central spike, and the positive skewness for the side-lobes and lowest frequencies (approximately log($f$/Hz) < -3.4) agree with reverse behaviour. In Fig.~\ref{profile_mean_rise-dec_compar} we directly compare the rising and decaying parts of the mean profile, and apparently the overall decaying part representing the lowest frequencies is faster. The shot profile behaviour does not contradict the \citet{scaringi2014b} finding then. Both the frequencies and skewness suggest that the authors finding based on the most dominant PDS feature does not refer to the central spike.

In order to compare the side-lobes location in the AGN and the CV, we estimated their relative positions using time scale ratios. The ratio of the rising time scale of the central spike $T_{\rm r}$ to the time coordinate of the rising side-lobe ($t_0$ in Eq.~\ref{lorentzian}) yields a value of $0.26 \pm 0.03$ for MV\,Lyr, while it is $0.31 \pm 0.05$ for the declining time scale $T_{\rm d}$ and location of the declining side-lobe. To get equivalent ratios in W2R\,1926+42 we performed a rough estimate of the side-lobes time locations from Fig.~5 in \citet{sasada2017}. The corresponding values are $-0.3 \pm 0.05$ and $0.3 \pm 0.05$\,d with uncertainty being half of the axis resolution of 0.1\,d. The ratios are $0.14 \pm 0.02$ and $0.20 \pm 0.03$ for the rising and declining part, respectively. When comparing these ratios of both objects we can conclude that they are different. Therefore, in spite of similarity the shot profile of W2R\,1926+42 is not just a time amplification of the MV\,Lyr shot profile.

\subsection{PDS structure}
\label{discussion_pds_structure}

MV\,Lyr was already studied in details yielding the discovery of a multicomponent PDS (\citealt{scaringi2012a}). Numerical modeling of the highest dominant component supports a sandwiched model origin, i.e. the fast variability with a frequency of log($f/{\rm Hz}) \simeq -3$ is generated by an inner evaporated hot corona (geometrically thick and optically thin disc) with a standard geometrically thin and optically thick accretion disc bellow (\citealt{scaringi2014a}). Such a corona radiates in X-rays, and this radiation is reprocessed by the geometrically thin accretion disc resulting in detected optical radiation. This was confirmed by the \xmm\ observation by \citet{dobrotka2017} yielding a presence of the log($f/{\rm Hz}) \simeq -3$ component in an X-rays PDS\footnote{\xmm\ PDS shows also the second break frequency at log($f/{\rm Hz}) \simeq -3.4$ confirming the reality of the feature.}. It suggests a very low mass accretion rate typical for dwarf novae in quiescence, which is not typical for a high state of a nova like rather resembling dwarf nova in outburst. This dilemma was explained by an evaporated corona having very low density yielding a low mass accretion rate with a standard geometrically thin disc below with a mass accretion rate typical for a nova like system in a high state.

Finally, the presence of the new high frequency PDS component at log($f$/Hz) = -2.42 in \xmm\ data suggests a coronal or boundary layer origin like at log($f$/Hz) = -3.0 and -3.4 mentioned above. This is an important conclusion because it rules out all other potential sources of the flickering like hot spot\footnote{Interaction of disc edge with the plasma stream from the secondary.}, outer disc or interaction of the overflowing stream from the secondary with the disc. However, this relates only to the radiation source. The initiating fluctuation can be generated elsewhere, even in an outer disc or a hot spot (see below in this section).

While X-ray detection localizes the radiation source, the localization of the fluctuation is not easy. The basic idea is, that every characteristic frequency or PDS component has its own origin in the disc. This hypothesis was studied by \citet{dobrotka2015} using simulations, yielding a complex model of the accretion flow from the more active outer disc rim, toward the central inner disc. If the studied fast variability would be a simple superposition of several different signals, they should be independent. Mainly fluctuations from outer parts of the disc would be indiscernible in the studied X-ray data. However, our finding from Section~\ref{pds_section} suggests a different concept based on single signal with a complex flare profile with substructures having their own characteristic time scales. Such a conclusion is clear from Fig.~\ref{pds_simul}, where at least two most dominant previously detected PDS components are present. If the flickering activity would be a superposition of different and independent signals, all patterns would be independent and not appearing simultaneously as seen in the averaged shot profile.

Furthermore, such a simple additive process is not real because of a typical rms-flux relation. This relation has a linear trend (\citealt{scaringi2012b} for MV\,Lyr) in general, which is typical for multiplicative processes. Therefore, the signals generated by different disc structures can not be independent and a correlation is expected. Following the propagating mass accretion fluctuation model the mass accretion variability at outer radii is propagating toward the center and is influencing the variability characteristics in inner regions. This means that a large mass accretion fluctuation at the outer disc propagates inward and generates further fluctuations on local viscous time scales which decrease inwards. All events with various time scales generated by this initial outer event must be correlated and yield an energy release at the boundary layer. This can be the reason why the complicated averaged shot profile has so much information with only part of it generated in the corona, and manifested as the highest frequency components. The rest can have the origin in different/outer disc parts. A definitive conclusion whether such a complex shot profile is a result of propagating accretion fluctuations is beyond the scope of this paper, and we leave the toy to other colleagues specialized in the modeling.

Finally, the presence of the several detected PDS components in simulated light curves using the observed profile supports the reality of these PDS components, because they are directly "seen" in the averaged shot profile.

\subsection{Shot profile evolution}

The shot profile evolution represented in Figs.~\ref{frekv_evol_spike} and \ref{frekv_evol_lobes} does not show anything worth discussion except the declining time scale $T_{\rm d}$ of the central spike. The data show more or less constant values within the errors except two values during the first flux plateau in the light curve around day 260, where a constantly rising trend experienced a short interval of more or less constant flux. The first idea is that this different $T_{\rm d}$ appears during the trend change, but there is a second plateau or trend change around day 460 where we do not see similar $T_{\rm d}$ deviation.

Another idea comes out, but it is just a speculation. The monotonic flux rise till the maximum around day 460 is slow. Such flux increase can be generated by a mass transfer increase from the secondary, resulting in increasing a mass accretion rate through the disc, and generating the flux rise. The subsequent flux decrease after day 460 can be explained by a similar reason, i.e. a mass transfer decrease. Such mass transfer variations are believed to be the origin of the long-term variability of nova like systems of VY\,Scu types (see \citealt{warner1995} for a review). However, if the first plateau is accompanied by the central spike $T_{\rm d}$ deviation, while the next plateau is not, this suggests a different origin of the two intervals with the constant fluxes. In this concept the variation in the mass transfer rate from the secondary generated the flux plateau around day 460, but an unknown fluctuation or instability appeared around day 260 and generated a temporal flux trend deviation.

Some other information concerning the time evolution of the shot profile can be taken from ACF calculations by \citet{kraicheva1999}. The authors used ground observations of MV\,Lyr obtained in 1992 and 1993. The side-lobes present in ACFs in Fig.~\ref{profile_evol} at approximately 700\,s from the zero (0.19\,h) are present in Fig.~4 of \citet{kraicheva1999} too, but the location is variable. This implies that the average shot profile possibly changes on time scales of days, i.e. on a shorter time scale than that of our samples shown in Fig.~\ref{profile_evol}.

\subsection{Comparison to W2R\,1926+42 and fluctuating mass accretion flow}

After our tests and the Monte Carlo simulations of \citet{sasada2017} we conclude that the detected shot profile and the similarities with the blazar W2R\,1926+42 are real. It is beyond the scope of this paper to offer a physical explanation of the detailed profile, but the very similar profiles suggest a common origin. The similarity is not only qualitative, but also quantitative. The time scale ratios $T_{\rm r}/T_{\rm d}$ in both objects are very comparable, but on the other side the side-lobes location relative to the central spike time scales are different. Both object are accretion powered but are very different in nature. This makes the similarities even more surprising.

\citet{sasada2017} first excluded geometrical effects in W2R\,1926+42 based on changes in viewing angle in a bent jet or gravitational lensing as the variability origin. The authors argued with the asymmetric central spike. In the case of MV\,Lyr we can exclude any gravitational lensing phenomena, and jet presence is questionable\footnote{Following thermal models, which describe jets generation by young stellar objects, the jets in CVs are not probable (\citealt{soker2004}). However, it was recently shown that CVs are significant radio emitters, and synchrotron radiation is the emission mechanism making the jets presence possible (\citealt{coppejans2015}, \citealt{coppejans2016}).}. Some further jet related rough cooling time scale and emission region size estimates by \citet{sasada2017} did not yield satisfying results. However, \citet{sasada2017} concluded that the rapid variations are plausibly explained by the particle-acceleration scenario in the jet. This is questionable and less probable in MV\,Lyr, because this CV has an outflow (\citealt{dobrotka2017}, \citealt{balman2014}), but it is more wind-like.

As a fast variability origin in MV\,Lyr \citet{scaringi2014a} proposed propagating accretion rate fluctuations in the disc. This unstable mass flow in the (central) disc is feeding also any (central) outflow. Therefore, it is easy to imagine that the source of variability is the fluctuating mass accretion rate unstably feeding all inner structures like the central corona (a radiation source in MV\,Lyr) or a jet (a radiation source in blazar). Within this scenario, every characteristics of the mass accretion rate fluctuation or unstable accretion in the (inner) disc should propagate into the corona or/and jet and modulate the radiation there, yielding similar radiation behaviour. However, following recent radio observations of NGC\,1275 by \citet{giovannini2018}, the jet does not need to be generated necessarily in the very central regions. The authors found a broad jet with a transverse radius larger than 250 gravitational radii at only 350 gravitational radii from the core. The jet can be launched from the disc then, and every mass accretion fluctuation can be naturally seen in the mass ejection.

Such interpretation needs further observational or theoretical studies. In the former case, the direct X-ray observations should say whether the optical blazar activity corresponds to the corona X-ray activity, while in the latter case, the same PDS modeling as made by \citet{scaringi2014a} should tests whether the observed blazar time scales are equivalent to the propagating mass accretion rate fluctuations.

However, some rough time scale estimates can give a clue, whether there is any connection between the observed flare durations. The basic time scales (dynamical $t_{\rm dyn}$, thermal $t_{\rm th}$ and viscous $t_{\rm visc}$) in the disc are connected as follows (see e.g. \citealt{king2008}):
\begin{equation}
t_{\rm dyn} \sim \alpha t_{\rm th} \sim \alpha (H/R)^2 t_{\rm visc},
\end{equation}
where $H$ and $R$ are the height scale of the disc and distance from the center, respectively. $t_{\rm dyn}$ is estimated from circular orbits:
\begin{equation}
t_{\rm dyn} \sim \left( \frac{R^3}{GM} \right)^{1/2},
\end{equation}
where $G$ and $M$ are the gravitational constant and white dwarf (black hole) mass, respectively. Finally, a realistic estimate for $t_{\rm visc}$ is
\begin{equation}
t_{\rm visc} \sim \frac{t_{\rm dyn}}{\alpha (H/R)^2}.
\label{eq_tvisc}
\end{equation}
Translating this into an equation for a radius $R$ estimate we get
\begin{equation}
R \sim \left( G\,M\,t_{\rm visc}^2\,[\alpha (H/R)^2]^2 \right)^{1/3}
\label{eq_r_corona}
\end{equation}

Following \citet{scaringi2014a} the PDS component at log($f/{\rm Hz}) \simeq -3$ is generated by an inner evaporated hot corona. Reanalysed X-ray data in this work suggest that also the central spike with a higher corresponding frequency of log($f/{\rm Hz}) \simeq -2.42$ in the PDS is generated in this corona, i.e. $t_{\rm visc} \simeq 263$\,s. The primary mass in MV\,Lyr is of about 0.73\,M$_{\rm \odot}$ (\citealt{hoard2004}) and $\alpha (H/R)^2 = 0.705$ (\citealt{scaringi2014a}). The resulting corona radius roughly estimated from Equation~(\ref{eq_r_corona}) is $1.5 \times 10^{10}$\,cm, which is close to the value of $0.8^{+0.2}_{-0.1} \times 10^{10}$\,cm derived by \citet{scaringi2014a}.

Performing the same calculus for the blazar with the black hole mass approximately $10^7$\,M$_{\rm \odot}$ (\citealt{mohan2016}) or $10^{7.8}$\,M$_{\rm \odot}$ (\citealt{marconi2003}), and the central spike time scale with the duration of 24547\,s (from \citealt{sasada2017}, log($f$/Hz) = -4.39), the equation~(\ref{eq_r_corona}) yields a value of $7.4 \times 10^{13}$ or $1.4 \times 10^{14}$\,cm for the corona radius\footnote{Supposing the same $\alpha (H/R)^2$ parameter as in MV\,Lyr for a geometrically thick corona.}. The typical corona in an AGN is of approximately 10 Swarzschild radii (\citealt{liu2017}), i.e. $10 \times 2MG/c^2 \simeq 3 \times 10^{13}$\,cm ($c$ being the speed of light) or $\simeq 2 \times 10^{14}$\,cm for both black hole mass estimates, respectively. Therefore, the values derived from the Equation~(\ref{eq_r_corona}) are comparable to 10 Swarzschild radii.

Both time scale estimates from the CV and the blazar are well explainable by the viscous processes in the central geometrically thick corona. If the estimated $t_{\rm visc}$ in the blazar is proportional to the plasma injection time into the jet, this could explain the above mentioned connection between the viscous processes in the corona and radiation fluctuations within the jet.

\subsection{Comparison to Cygnus\,X-1 and aperiodic mass accretion model}

\kepler\ data are not the only data suitable for the superposed shot profile study. The same was made by \citet{negoro2001} in the case of \ginga\ data of Cyg\,X-1. The shot profile of this XRB in X-rays has a similar shape as the two optical shot profiles discussed in this paper, i.e. a central spike with two humps, one before (at -1.85\,s from zero) and one after (at 0.96\,s from zero) the spike. The time scales are much shorter than those of the other two systems with the central spike having durations of the order of 0.1 - 1\,s (see Table~1 or Fig.~1 in \citealt{negoro2001}). This time scales were estimated by two superposed exponential functions, with the shortest time scale (center of the spike) rising more gradually, but with the longer time scale behaving in the same way as in the case of MV\,Lyr and W2R\,1926+42, i.e. having a declining part of the central spike more gradual with $T_{\rm r}/T_{\rm d} = 0.72/1.13 = 0.64 \pm 0.02$. This value is comparable to both previously discussed objects.

The variability profile of Cyg\,X-1 can be derived from skewness measurement like in the MV\,Lyr case described in Section~\ref{shot_kepler}. \citet{maccarone2002} and \citet{scaringi2014b} analysed \rxte\ data of this XRB, and found skewness behaviour comparable to the shot profile shape in \ginga\ data, i.e. time scales from approximately 1\,s have negative skewness which agrees with the above derived $T_{\rm r}/T_{\rm d}$ ratio (slower decay)\footnote{The skewness resolution is not enough for the shorter time scales of $\sim 0.1$\,s studied by \citet{negoro2001}.}. The skewness returns back to positive values for larger time scales of approximately 5\,s (comparable to largest 0.15\,Hz sine function in Fig.~1 of \citealt{negoro2001}) describing the overall rising/decaying trend, which makes the profile behaviour similar to MV\,Lyr. 

Using the same $T_{\rm r}$ and $T_{\rm d}$ values and side-lobes locations at $-1.85 \pm 0.01$ and $0.96 \pm 0.03$\,s for the rising and declining side-lobe, respectively, we can estimate the relative side-lobes location represented by the time scale ratios as in Section~\ref{shot_kepler}. The ratios are $0.39 \pm 0.01$ and $1.18 \pm 0.05$ for the rising and declining part, respectively. taking into account that the side-lobes locations were not stable and changed from $-4$\,s to $-1.85$\,s in \ginga 1987 and 1990 observations, respectively (\citealt{negoro1995}), the ratio in the rising part is comparable to the MV\,Lyr value of $0.26 \pm 0.03$.

All three discussed objects are accretion powered, and \citet{negoro1995} proposed a physical interpretation based on the unstable mass accretion rate within the disc, the so called aperiodic mass accretion model. Following this model, clumps of accreting matter drift inwards releasing the gravitational energy in the form of a flickering flare. \citet{negoro1995} proposed the shape of the flare rising branch to be in the form of
\begin{equation}
F(t) = \frac{A}{(\tau - t)^{\alpha}},
\end {equation}
where A, $\tau$ and $\alpha$ are constants. The most important and sensible parameter being the power $\alpha$. The author fitted this function to superposed profiles of Cyg\,X-1 taken by \ginga\ in 1987 and 1990 and got $\alpha = 0.7$ in both cases. We performed the same procedure in the case of MV\,Lyr with corresponding fits shown as red lines in Fig.~\ref{profile_mean}. In the $N_{\rm pts} = 200$ case the fit did not describe the profile well, therefore we excluded the two humps from the fitting process (between -2550 and -400, the interval is marked by two arrows in bottom panel of Fig.~\ref{profile_mean}). Both fits yield $\alpha = 0.7$ as in Cyg\,X-1 \ginga\ data.

The similarity supports the same physical process. We talk about the aperiodic mass accretion in Cyg\,X-1 and the propagating mass accretion fluctuations in MV\,Lyr. Such mass accretion fluctuation within the central corona can be imagined as a clump of accreting matter drifting inwards and releasing gravitational energy.

\subsection{Possible contamination and other detections}

\citet{borisov1992} and \citet{skillman1995} studied ground data of MV\,Lyr and concluded possible presence of a superhump period of 0.1379\,d with a corresponding frequency of log($f$/Hz) = -4.08, while the latter authors concluded that their data were less strongly supportive. The search for such coherent periodicity is beyond the scope of this paper, but we performed very rough attempt to search for this signal and did not succeed. However, such a frequency is much lower than the highest frequency components of log($f$/Hz) = -3.0 and -2.4 representing the side-lobes and the central spike which we study here in details. Therefore, any contamination by a superhump (if present) is irrelevant.

\citet{dobrotka2016} studied the rms-flux relation of two SU\,UMa systems observed by \kepler. While V1504\,Cyg shows the typical linear rms-flux relation, V344\,Lyr does not. However, the latter exhibit superhump activity and if the flickering has a linear rms-flux relation, the additional superhump flux must deform this linearity. The authors showed by simple simulations how superhump activity with varying amplitude deformed the linearity, and got a very comparable rms-flux relation to the observed data. This implies, that if any significant superhump activity is present in MV\,Lyr, the rms-flux relation must differ from linearity. But this is not the case (\citealt{scaringi2012b}). If still present, it must be of very small amplitude or significance. \citet{borisov1992} reported a 4\% amplitude.

Finally, \citet{borisov1992} and \citet{skillman1995} together with \citet{kraicheva1999} found a QPO at a period near 47 minutes. The corresponding frequency of log($f$/Hz) = -3.45 is in perfect agreement with one PDS component detected by \citet{scaringi2012a} in the optical \kepler\ data and by \citet{dobrotka2017} in the X-ray \xmm\ light curve. This power excess is also seen in the simulations (Fig.~\ref{pds_simul}) based on the averaged shot profile method in this paper. Therefore, all findings are consistent.

\subsection{Method}

Finally, we note that the averaged shot profile method is not a generally accepted method. However, it brings a new way how to investigate the fast variability. It has some advantages compared to the standardly used PDS or ACF methods.

First is the fact, that the shot profile method use only fragments of the light curve. If the calculation of the ACF requires evenly spaced data, this can be problematic if the light curve has many interruptions. Interpolation of the data by anything is already bringing false information and can affect the results. The shot profile method uses only the flares, i.e. short fragments of the data.

Another advantage is seen when comparing both panels of Fig.~\ref{profile_evol}. The left panel shows that the central spike is rising in amplitude with the average overall flux, while the side-lobe amplitudes are more or less stable. This behaviour not seen in ACF suggests that the central spike (the highest frequencies) is responsible for the rms-flux relation detected by \citet{boeva2011} in ground observations and by \citet{scaringi2012b} using the \kepler\ data.

Furthermore, some PDS components are directly "seen" in the shot profile. All the Fourier based methods are somehow abstract descriptions of the reality. A shot profiles directly show the phenomenon, i.e. components location, variability amplitudes, any asymmetry etc. Mainly the latter is important as it is not seen in the ACF. As a consequence we detected similarity of the variability in MV\,Lyr and the blazar. Such similarity is hardly deducible from the PDSs. Finally, individual PDS components can have similar frequencies and be non-distinguishable in the PDS, while they are seen as separate features in the shot profile.

Moreover, the detection of the new high frequency component was done thanks to the averaged shot profile and subsequent light curve/PDS simulations. Without the small power deviation of two PDS points based on the simulations we would not be motivated to search for the component in the real data. The advantage is that such a PDS is based on artificial light curves constructed with the averaged profile, where the whole noise is smoothed by the averaging process, and only real features remain. Finally, this positive result certifies the harmlessness of the inappropriate shot noise model.

\section{Summary and conclusions}

The results of this work can be summarized as follows:

(i) The superposed averaged shot profile of flickering activity in MV\,Lyr \kepler\ data shows a complex structure with a central spike and side-lobes on both sides of the central spike. These various substructures correspond to a single dominant component in the PDS detected by \citealt{scaringi2012a}. The standard PDS is not able to distinguish them. We confirmed the reality of the features using a shot noise model. Such approach is purely phenomenological, because it does not fulfill all details of the underlying physics. Moreover, additional tests showed that non evenly sampled \kepler\ data due to barycentric correction has no effect on the results. Finally, the credibility of the shots is supported by time scale ratios of the rising and declining shot parts when compared to skewness measurement of \citet{scaringi2014b}.

(ii) Time evolution of the shots in MV\,Lyr does not show any significant evolution or correlation with the flux except during a short flux plateau. The declining time scale of the central spike increased during this time interval.

(iii) A very similar complex shot profile structure appears also in \kepler\ data of the blazar W2R\,1926+42 (\citealt{sasada2017}) and \ginga\ data of Cyg\,X-1 (\citealt{negoro1994,negoro2001}).

(iv) The time scales of the rising and declining branches of the central spike have very comparable ratios in all three very different objects in nature.

(v) The rise of the shots in MV\,Lyr and Cyg\,X-1 follows a trend with the same power law (see \citealt{negoro1995} for Cyg\,X-1 case).

(vi) The similarities between the averaged shot profiles in the three very different objects in nature but powered by accretion imply a similar physical origin of the variability within the accretion process. The origin can be the same (propagating accretion rate fluctuations in the disc feeding the inner disc), but emitting regions can be different (corona in MV\,Lyr and jet in W2R\,1926+42).

(vii) Simple shot noise simulations using the averaged shot profile yield to the identification of another high frequency PDS component not identified so far. The component is noticeable with low significance in \kepler\ data but quite clear in \xmm\ EPIC/pn light curve. The latter suggests that the radiation source is in the inner region of the disc, i.e. the corona or boundary layer.

\section*{Acknowledgement}

AD was supported by the ERDF - Research and Development Operational Programme under the project "University Scientific Park Campus MTF STU - CAMBO" ITMS: 26220220179. HN was partially supported by Grants-in-Aid for Scientific Research 16K05301 from the Ministry of Education, Culture, Sports,  Science and Technology (MEXT) of Japan.

\bibliographystyle{aa}
\bibliography{mybib}

\label{lastpage}

\end{document}